# Thermal Behavior of a Single Magnetic Vortex Studied with Magnetotransport

Sergi Lendinez, Junjia Ding, Tomas Polakovic, John E. Pearson, Axel Hoffmann, and Valentyn Novosad*

*Materials Science Division, Argonne National Laboratory, Argonne IL, 60622, USA*

*novosad@anl.gov

Spin textures such as skyrmions and magnetic vortices are good candidates for a variety of applications, such as magnetic memories, oscillators and neuromorphic computing. Understanding the magnetic process of these systems is important, as it determines the system's response in field and frequency. In this work, we investigated the magnetization process of single microdisks by measuring their magnetotransport properties as a function of temperature. The strong dependence of resistance on the disks magnetic state helped us understand the magnetization configurations of a single microdisk for different temperatures and fields. We determined the thermal barriers for the nucleation and annihilation processes by fitting the nucleation and annihilation fields to an exponential model. Moreover, we observed and characterized the domain wall depinning effect for temperatures below 100 K. This effect prevents the formation of a magnetic vortex during the nucleation process.

## I. INTRODUCTION

Magnetic vortices can be the ground state of certain ferromagnetic systems and have been the subject of study for the past decade for potential applications[1] ranging from elements of random access magnetic memories[2,3] to cancer cell destruction[4]. Magnetic vortices can be observed in soft ferromagnetic disks with micron and sub-micron diameters. The geometry constraints the magnetization to curl around the center[5], with the center spins pointing out of plane due to the strong exchange interaction between antiparallel spins[6]. The nucleation, annihilation and displacement processes of the magnetic vortices have been extensively investigated, and several groups have reported interesting phenomena[7–15]. One particular interesting topic is the effect of defects on the magnetic properties of nanostructures. It has been observed that the defects and pinning sites play an important role in the domain wall motion of nanowires[16,17] and in the gyrotropic frequency of the vortex core[18] in nanodisks. In another study, the coercive field and the residual magnetization of magnetic disks changed as a function of temperature due to the pinning potentials on the samples[9]. Moreover, two different mechanisms of the magnetization evolution were proposed at different temperature regimes: at low temperatures, nucleation and annihilation were governed by jumping over an energy barrier, while high temperature processes were dominated by the decrease in magnetization[14]. So far, a detailed study of domain wall motion at low temperatures in single magnetic disks with a magnetic vortex configuration remains missing.

In this work, we used electrical contacts on top of magnetic disks to measure the change in resistance of a single magnetic disk at different temperatures from 2 K to 300 K with an in-plane magnetic field. The results at different temperatures were compared with micromagnetic simulations and a similar trend observed in both measurements and simulation. The micromagnetic simulations also revealed a better picture of the magnetization process as a function of temperature. Finally, we analyzed different energy barriers involved in the magnetization process of the magnetic disk.

## II. EXPERIMENT

The sample was fabricated using a multistep electron-beam lithography (EBL) process. First, the disk with 1 um diameter and alignment marks were defined on a positive bilayer resist of ZEP520A and PMGI SF2 on a silicon substrate, accompanied by e-beam evaporation and lift-off process of a 50 nm-thick $Ni_{80}Fe_{20}$ film. The second step was an EBL patterning of 10-nm-wide nanocontacts, followed by deposition of 5 nm Ti and 100 nm Au and lift-off, completing the fabrication process. . Scanning electron microscopy was used to confirm alignment after the lithography steps [Fig. 1(a)].

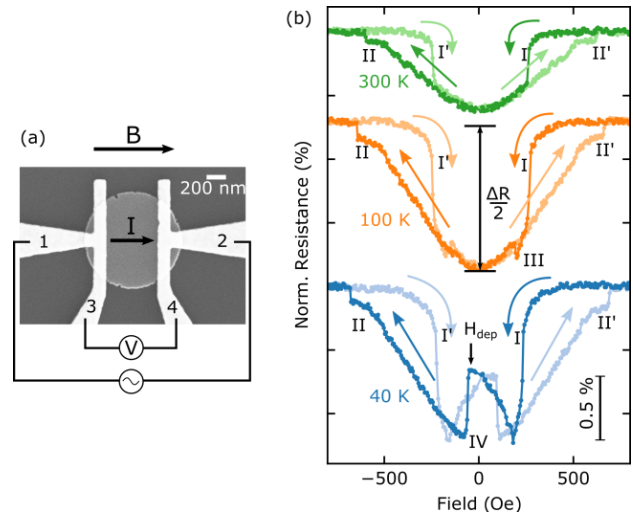

FIG. 1. (a) SEM image of a 1 um-diameter Py disk and Ti/Au nanocontacts to measure magnetotransport. The current was applied through the external contacts (1 and 2), and the voltage response is measured on the middle contacts (3 and 4) with a lock-in technique. (b) AMR curves of the disk passing a 10 $\mu$A amplitude current at 300 K (top, green), 100 K (middle, orange), and 40 K (bottom, blue). Dark colors show the sweep down, and light colors show the sweep up, as indicated by the arrows.

The electrical response of the disk was measured with 4-probe electrical contacts. Referring to the disk shown in Fig. 1(a), a

1.4 kHz sine-modulated current of 100 μA in amplitude was applied to the disk through ports 1 and 4, and the voltage response was picked up by a lock-in amplifier through ports 2 and 3. The signal gain was 500, achieved by a low-noise pre-amplifier. Our disks had a typical resistance value in the order of 1-10 Ω, which corresponds to a resistivity value of $A = \rho \sim 50\ \mu\Omega \cdot cm$, in good agreement with known Py resistivity[19]. The resistance was measured as a function of an in-plane magnetic field parallel to the current in the range from -1000 Oe to +1000 Oe for different temperatures from 2 K to 300 K. Micromagnetic simulations on a disk of 1 um diameter and 50 nm thickness were carried out using the LLG micromagnetics simulator[20], with a cell size of 4 x 4 x 50 nm³. Standard parameters for Py (saturation magnetization $M_S = 800\ \text{emu/cm}^3$, exchange stiffness constant $A = 1.05\ \mu\text{erg/cm}^3$) were used in the simulation. The damping parameter and the convergence criterion were changed in different simulations.

### III. RESULTS AND DISCUSSION

The change in resistance measured in this work was produced by the anisotropic magnetoresistance (AMR) of the device. In AMR, the resistivity depends on the relative orientation between the magnetization and the current: $\rho = \rho_\perp + (\rho_\parallel - \rho_\perp)(\hat{\jmath} \cdot \boldsymbol{m})^2 = \rho_\perp + (\rho_\parallel - \rho_\perp)\cos^2 \alpha$, with $\rho_\perp$ the perpendicular resistivity, $\rho_\parallel$ the longitudinal resistivity, $\hat{\jmath}$ the current density vector, $\boldsymbol{m}$ the unit vector pointing in the magnetization direction, and α the angle between $\hat{\jmath}$ and $\boldsymbol{m}$. In other words, the highest resistance values can be observed when the magnetization is parallel to the current direction ($\alpha = 0°$), while the lowest resistance values are obtained when the two directions are perpendicular with each other ($\alpha = 90°$). A maximum 2% of the total resistance difference for $\alpha = 0°$ and $\alpha = 90°$ was found in our device at 300 K (with an applied field fixed at 1000 Oe).

Figure 1(b) shows the AMR curves of the disk for $\alpha = 0°$ at three representative temperatures: 300 K (top, green), 100 K (middle, orange), and 40 K (bottom, blue). The curves have been shifted for clarity. At 300 K, the change in resistance from 0 Oe to 1000 Oe is in the range of 1%, corresponding to half of the maximum resistance change (2%) [Fig. 1(b)]. This suggests that the same amount of spins point parallel and perpendicular to the current at remanence. Interestingly, the measured curves overlap with each other in the forward and backward half of the loop. Both phenomena agree with the behavior of a vortex state in the disk. Moreover, nucleation (annihilation) of the vortex core can be identified as a step-like resistance change in the AMR curve for the forward half of the measurement around 200 (-600) Oe, indicated by I (II) in Fig. 1(b). Symmetric steps at -200 Oe [point I' in Fig. 1(b)] and 600 Oe [point II' in Fig. 1(b)] were observed in the backward half of the measurement. All these features indicate a regular vortex-nucleation-annihilation process in the disk[21–23].

An additional resistance drop was observed around 200 Oe [point III in Fig. 1(b)] in the result for 100 K temperature. The resistance then jumped back up after a few Oe to the reversible linear behavior characteristic of a magnetic vortex, similar to the results at 300 K. Interestingly, the shape of the AMR curve changed more drastically when the temperature was reduced to 40 K [Fig. 1(b)]. After the nucleation (the step-like decrease in resistance indicated by I) around 200 Oe, the resistance rose again as the field decreased. This counter-intuitive behavior indicates that more spins align with the current as the magnetic field is further decreased. The resistance jumped back to the reversible curve corresponding to a magnetic vortex [point IV in Fig. 1(b)] when a reversal field of $H_{\text{dep}} = -50$ Oe was applied [indicated in Fig. 1(b)].

In order to understand the magnetic configurations of the system at different temperatures, we performed micromagnetic simulations. Figure 2 shows the simulated AMR curves with different simulation parameters. Here, a small sensing current of 1 pA was applied in the same direction as the magnetic field. Good qualitative agreement between the simulated and experimental results can be observed by comparing Fig. 2 to Fig. 1(b). The top green curve shows a similar behavior to the experimental response at 300 K shown in Fig. 1. With an initial magnetic field of 1000 Oe, the disk was first in a saturated state (1). When the magnetic field decreased, the magnetization started curling into a C-shape (2a), and the resistance dropped accordingly. By further decreasing the magnetic field, the magnetic vortex state (3) was created at the nucleation field (point II), and the resistance dropped again. As the field approached 0, the vortex moved to the center of the disk, and the resistance reached the minimum value. When a reversed field was applied, the resistance increased again as the vortex moved perpendicularly to the magnetic field, until the magnetic field reached the annihilation field (point II)). The single domain state formed with a sharp increase of the resistance as the reversed field is increased. The damping parameter is 0.01 and the convergence is $10^{-6}$ in this simulation.

The orange, middle curve in Fig. 2 reproduces the peak after nucleation in the AMR curve measured at 100 K shown in Fig. 1(b) (point III). In this case, the damping parameter was 1 and the convergence was kept at $10^{-6}$. It can be observed that the magnetization goes from the saturated state (1) to an S-shape configuration (2b). When the magnetic field was further decreased, the S-shape eventually collapsed into one single vortex (3) with spin wave emission around 220 Oe. Since the S-shape configuration is characterized by two vortices on opposite edges of the disk with opposite chiralities and a central part with the magnetization perpendicular to the magnetic field, the resistance slightly increased when the magnetic vortex was nucleated, as there are less spins perpendicular to the field, and a peak was observed (point III).

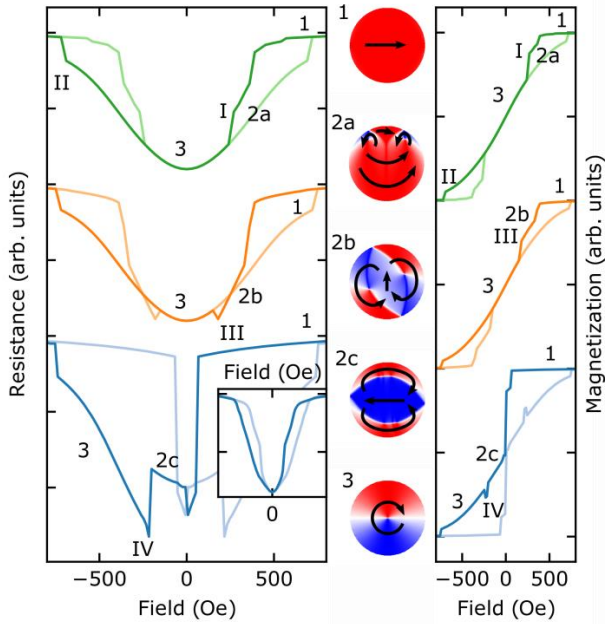

FIG. 2. AMR (left) and magnetization (right) curves obtained from simulations with different parameters, reproducing the trends of the experimental AMR curves at 300 K (top, green), 100 K (middle, orange), and 40 K (bottom, blue). The middle row displays magnetic configurations at different magnetic fields: (1) saturated state, (2a) C-shape configuration, (2b) S-shape configuration with two vortices, (2c) an intermediate state with a large longitudinal magnetization, and (3) a magnetic vortex.

The general trend of the 40 K AMR curve of Fig. 1(b) is reproduced by the blue, bottom curve of Fig. 2 in which the convergence criterion has been increased to $5 \times 10^{-5}$. In this simulation, an intermediate state with a central domain parallel to the magnetic field (2c), which increases the resistance, formed prior to the vortex state. This intermediate state showed a higher resistance than the vortex state, as observed experimentally for the 40 K curve in Fig. 1(b). This simulation result was obtained by further increasing the convergence criterion to $5 \times 10^{-5}$. A smaller convergence value in the simulation can be considered as an effective method to simulate room temperature experiments. It has the effect of letting the magnetization evolve for a longer time, hence reaching a state closer to equilibrium and the magnetization overcomes energy barriers thanks to thermal activation, which is more precisely simulated by a closer-to-equilibrium state. On the contrary, at low temperatures the magnetization cannot overcome some barriers and hence the magnetization may remain in a metastable state. Higher convergence criteria produce magnetization states farther from equilibrium. In the inset of Fig. 2 we present a simulation obtained with the same simulations parameters as the blue curve, but at a temperature of 100 K. It shows that the magnetic process changes to one similar to a lower convergence criterion.

Figure 2 also shows the corresponding simulated magnetization curves obtained during the field sweeps, at the same time as the resistance values were computed. The results show that the AMR technique was more suitable to distinguish some of the intermediate states. For example, the magnetization curves of the top, green curve and the middle orange curve were similar, and could only be distinguished by the particular values on the nucleation and the annihilation fields, making the C-state (2a) and the S-state (2b) nearly indistinguishable. On the other side, the corresponding resistance curves show a clearly distinct behavior, with the appearance of a peak in the case of an S-shape configuration. Moreover, by looking at the bottom, blue magnetization curve it would not be possible to observe that the magnetization is in a different state at low magnetic fields (2c), and the characteristic feature in the magnetization curve when the state changes to the magnetic vortex (point IV) is relatively much smaller than in the case of the resistance.

In order to further elucidate the importance of temperature in the system, we studied the different energies involved in the magnetization process of the system, i.e., thermal excitation of spin waves following Bloch's law, magnetization reversal with a thermal energy barrier following Arrhenius's law, and depinning of a domain wall.

First, we studied the effect of temperature on the change in the total AMR. The change in resistance between the saturated state at $\alpha = 0°$ and $\alpha = 90°$ ($\Delta R$ shown in Fig. 1) was much lower at room temperature than that at lower temperatures. Figure 3(a) shows the temperature dependence of $\Delta R/\Delta R_0$, where $\Delta R_0$ is the value at the lowest measured temperature. The data were fitted to the expression $\Delta R = a - bT^{3/2}$. There is a good agreement between the fitting results [solid line in Fig. 3(a)] and the measurements, with a value of $b_R = (6.5 \pm 0.4) \times 10^{-5} \, K^{-3/2}$. The inset of Fig. 3(a) shows the magnetization of a reference 50-nm-thick Py film, normalized to the value at the lowest temperature $M_0$. We observed that the effects on the magnetization are much smaller than in the case of the resistance, but it also followed a $T^{3/2}$ law. The fitting to the same expression gave a value of $b_M = (7.3 \pm 0.1) \times 10^{-6} \, K^{-3/2}$, one order of magnitude smaller, which is attributed to thermal excitation of spin waves.

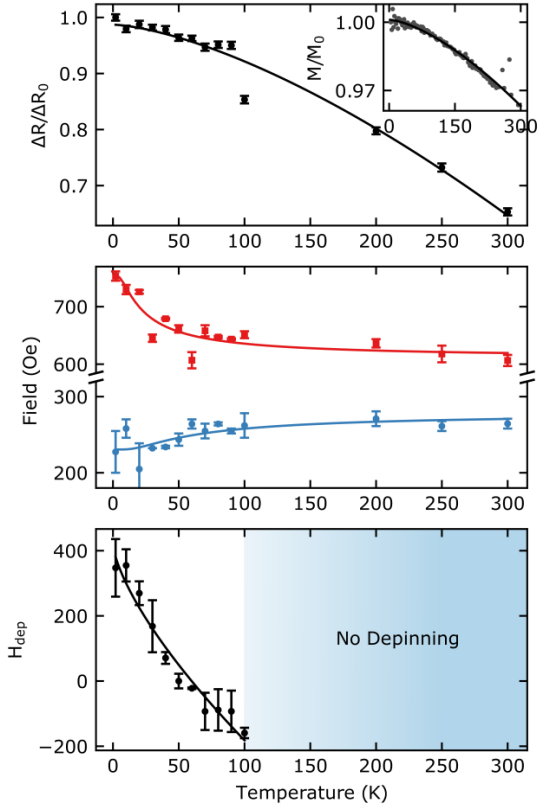

FIG. 3. Temperature dependence of (a) the total change in resistance, (b) nucleation and annihilation fields, and (c) the depinning field. The lines show fits as described in the text: $T^{3/2}$ Bloch's law in (a), a thermal barrier in (b), and domain wall depinning in (c). The inset in (a) shows the evolution of the magnetization of a reference Py.

Second, the energy barrier for nucleation and annihilation was investigated by measuring $H_n$ and $H_{an}$ as a function of temperature [Fig. 3(b)]. The average value between the loops with decreasing and increasing field was used to reduce the uncertainty, and the positive value is presented. The data were fitted to an exponential function, corresponding to the system overcoming a thermal energy barrier (Arrhenius's law): $H_{(n,an)} = H'_{(n,an)} \exp(-U/k_B T) + H_{(n,an)0}$. A value of $U_{an} = 20 \pm 7$ K was obtained for the annihilation field, and $U_n = 60 \pm 40$ K for the nucleation field. The different barriers for the nucleation and annihilation fields indicate that the reversal processes are different for each of these mechanisms. This result is different than those obtained in previous temperature-dependent studies of the characteristic fields, where they obtained different trends for high and low temperature regimes[14]. These differences could be produced by the different size, shape and defects of the samples. However, Ščepka et al.[24] found a similar descending trend in the nucleation fields as the one observed in our measurements.

Moreover, the intermediate states (S-shape for 100 K, and a state with an extra longitudinal magnetization for 40 K) before the nucleation of a vortex state at low temperature [orange, middle curve and blue, bottom curve in Figs. 1(b) and 2] can be considered as a good indicator of the domain wall depinning energy barrier. In Fig. 3(c) we plot the field at which the resistance jumped from the intermediate state to the resistance corresponding to the vortex state, averaged for the forward and backward sweeps and taking the positive value, $H_{\text{dep}}$. This jump in resistance is similar to previous observations on domain walls moving through a magnetic nanowire[16], and here we analyzed the corresponding barrier of domain wall depinning in our disk. The model was derived from switching of the magnetization via thermal activation over a field-dependent energy barrier[25–29]. In the case that the energy barrier depends with field as $U \sim (1 - H/H_0)^{3/2}$, the expression of the depinning field as a function of temperature can be written as[27]

$$H_{\text{dep}} = H_0 \left\{ 1 - \left[ \frac{k_B T}{E_0} \ln \left( \frac{\Gamma_0 k_B H_0 T}{1.5 E_0 r \epsilon} \right) \right]^{2/3} \right\}$$

where $H_0$ is the depinning field at 0 K, $\epsilon = \sqrt{1 - H/H_0}$, $k_B$ is the Boltzmann constant, $\Gamma_0$ is the attempt frequency, $r = 10$ Oe/s the field sweep rate, and $E_0$ is the energy barrier for a domain wall depinning. We neglect the term with $\epsilon$, as its effect is small. We considered a value of $\Gamma_0 = 10^8$ Hz, and obtained a depinning energy barrier for the domain wall of $E_0/k_B = 1130 \pm 50$ K and a depinning field at 0 K of $H_0 = 420 \pm 30$ Oe. The barriers obtained in previous studies with similar steps in resistance for domain walls moving over a defect in nanowires were much higher[16,17].

While other samples with the same and different sizes gave the same general trends described in this study, the shape of the AMR curves and exact energy values varied from sample to sample. This is a further indication that pinning is playing an important role in the magnetic configuration at low temperatures, as the defects and pinning sites will differ between samples.

**IV. CONCLUSIONS**

In summary, we studied the temperature dependence of the magnetization processes of a single magnetic disk using magnetotransport measurements. Our measurements show the importance of measuring single disks. It allowed us to capture the details of the magnetization process, which is not possible for multiple-disk measurements. We found different temperature-dependent phenomena for higher and lower temperature regions. We confirmed the thermal excitation of spin waves in the disk, indicated by the decrease of the total AMR following a $T^{3/2}$ law. We have found different thermal energy barriers for the nucleation and annihilation processes. Moreover, by reaching temperatures below 100 K we were able to observe a different regime in the magnetic configuration, governed by pinning, and we have characterized the domain wall depinning in our system. The pinning in our samples had the effect of changing the remanent state at 0 field. This finding shows the importance of measuring single disks, as a non-zero remanent magnetization--such as the one observed by Shima et al.[9]--might be produced not only by the pinning of the core out of the center, but also by a different state. The determination of the exact magnetic configuration of these states at low temperatures requires further study.

ACKNOWLEDGMENTS

Work at Argonne, including use of the Center for Nanoscale Materials, an Office of Science User Facility, was supported by the U. S. Department of Energy, Office of Science, Basic Energy Sciences, under Contract No. DE-AC02-06CH11357.

REFERENCES

[1] R.P. Cowburn, J. Magn. Magn. Mater. **242–245**, 505 (2002).

[2] S. Kasai, K. Nakano, K. Kondou, N. Ohshima, K. Kobayashi, and T. Ono, Appl. Phys. Express **1**, 91302 (2008).

[3] A. Drews, B. Krüger, G. Meier, S. Bohlens, L. Bocklage, T. Matsuyama, and M. Bolte, Appl. Phys. Lett. **94**, 62504 (2009).

[4] D.-H. Kim, E.A. Rozhkova, I. V. Ulasov, S.D. Bader, T. Rajh, M.S. Lesniak, and V. Novosad, Nat. Mater. **9**, 165 (2010).

[5] R.P. Cowburn, D.K. Koltsov, A.O. Adeyeye, and M.E. Welland, Phys. Rev. Lett. **83**, 1042 (1999).

[6] T. Shinjo, T. Okuno, R. Hassdorf, † K. Shigeto, and T. Ono, Science (80-. ). **289**, 930 (2000).

[7] V. Novosad, K.Y. Guslienko, H. Shima, Y. Otani, K. Fukamichi, N. Kikuchi, O. Kitakami, and Y. Shimada, IEEE Trans. Magn. **37**, 2088 (2001).

[8] V. Novosad, M. Grimsditch, K.Y. Guslienko, P. Vavassori, Y. Otani, and S.D. Bader, Phys. Rev. B **66**, 52407 (2002).

[9] H. Shima, V. Novosad, Y. Otani, K. Fukamichi, N. Kikuchi, O. Kitakamai, and Y. Shimada, J. Appl. Phys. **92**, 1473 (2002).

[10] S.B. Choe, Y. Acremann, A. Scholl, A. Bauer, A. Doran, J. Stöhr, and H.A. Padmore, Science **304**, 420 (2004).

[11] V. Novosad, F.Y. Fradin, P.E. Roy, K.S. Buchanan, K.Y. Guslienko, and S.D. Bader, Phys. Rev. B **72**, 24455 (2005).

[12] K.S. Buchanan, P.E. Roy, M. Grimsditch, F.Y. Fradin, K.Y. Guslienko, S.D. Bader, and V. Novosad, Nat. Phys. **1**, 172 (2005).

[13] J.-G. Caputo, Y. Gaididei, F. Mertens, and D. Sheka, Phys. Rev. Lett. **98**, 56604 (2007).

[14] G. Mihajlović, M.S. Patrick, J.E. Pearson, V. Novosad, S.D. Bader, M. Field, G.J. Sullivan, and A. Hoffmann, Appl. Phys. Lett. **96**, 112501 (2010).

[15] J. Ding, G.N. Kakazei, X. Liu, K.Y. Guslienko, and A.O. Adeyeye, Sci. Rep. **4**, 4796 (2014).

[16] A. Himeno, T. Okuno, T. Ono, K. Mibu, S. Nasu, and T. Shinjo, in *J. Magn. Magn. Mater.* (North-Holland, 2005), pp. 167–170.

[17] Y. Gao, B. You, H.L. Yang, Q.F. Zhan, Z. Li, N. Lei, W.S. Zhao, J. Wu, H.Q. Tu, J. Wang, L.J. Wei, W. Zhang, Y.B. Xu, and J. Du, AIP Adv. **6**, 125124 (2016).

[18] R. Compton and P. Crowell, Phys. Rev. Lett. **97**, (2006).

[19] A.F. Mayadas, J.F. Janak, and A. Gangulee, J. Appl. Phys. **45**, 2780 (1974).

[20] M.R. Scheinfein, Http://llgmicro.home.mindspring.com (1997).

[21] S. Kasai, Y. Nakatani, K. Kobayashi, H. Kohno, and T. Ono, Phys. Rev. Lett. **97**, 107204 (2006).

[22] M. Sushruth, J.P. Fried, A. Anane, S. Xavier, C. Deranlot, M. Kostylev, V. Cros, and P.J. Metaxas, Phys. Rev. B **94**, 100402 (2016).

[23] X. Cui, S. Hu, M. Hidegara, S. Yakata, and T. Kimura, Sci. Rep. **5**, 17922 (2015).

[24] T. Ščepka, T. Polakovič, J. Šoltýs, J. Tóbik, M. Kulich, R. Kúdela, J. Dérer, and V. Cambel, AIP Adv. **5**, 117205 (2015).

[25] L. Gunther and B. Barbara, Phys. Rev. B **49**, 3926 (1994).

[26] W. Wernsdorfer, K. Hasselbach, A. Benoit, B. Barbara, B. Doudin, J. Meier, J.-P. Ansermet, and D. Mailly, Phys. Rev. B **55**, 11552 (1997).

[27] J.G. Lok, A. Geim, U. Wyder, J. Maan, and S. Dubonos, J. Magn. Magn. Mater. **204**, 159 (1999).

[28] D.B. Gopman, D. Bedau, G. Wolf, S. Mangin, E.E. Fullerton, J.A. Katine, and A.D. Kent, Phys. Rev. B - Condens. Matter Mater. Phys. **88**, 100401 (2013).

[29] J.A.J. Burgess, J.E. Losby, and M.R. Freeman, J. Magn. Magn. Mater. **361**, 140 (2014).